# A Component Based Approach to Scientific Workflow Management

J.-M. Le Goff, Z. Kovacs

*CERN, Geneva, Switzerland*

N. Baker, P. Brooks, R. McClatchey

*Centre for Complex Cooperative Systems, Univ. West of England, Frenchay, Bristol BS16 1QY UK*

*Abstract*

CRISTAL is a distributed scientific workflow system used in the manufacturing and production phases of HEP experiment construction at CERN. The CRISTAL project has studied the use of a description driven approach, using meta-modelling techniques, to manage the evolving needs of a large physics community. Interest from such diverse communities as bio-informatics and manufacturing has motivated the CRISTAL team to re-engineer the system to customize functionality according to end user requirements but maximize software reuse in the process. The next generation CRISTAL vision is to build a generic component architecture from which a complete software product line can be generated according to the particular needs of the target enterprise. This paper discusses the issues of adopting a component product line based approach and our experiences of software reuse.

# 1. Introduction

As component technology gradually evolves and matures so system developers will gradually migrate from systems composed of interoperable objects to those composed of interoperable components. One of the main motivations for this migration is the potential of software reuse and its associated benefits of cost reduction and time to market of software products. Component-based software development is concerned with constructing software artifacts by assembling prefabricated configurable building blocks. However software reuse is concerned with more than binary components. For many organizations it is the generation and application of generic software assets that are reusable across a family of target products. Binary components are just one view of the software development process. The creation and evolution of graphical models to visualize specific aspects of software artifacts is another view. What is required is some software development process that couples these high-level development approaches with implementation approaches. This paper opens with a brief discussion of the context and motivations for this research followed by an outline of software product lines. The issues and the team's experience of software reuse are discussed. The final part of the paper concentrates on the divide between object based modeling and component based development, which is preventing software reuse from reaching its full potential.

# 2. Motivation

CRISTAL is a scientific workflow system[1] that is being used to control the production and assembly process of the CMS Electromagnetic Calorimeter (ECAL) detector at CERN Geneva. Detector production is a collaborative effort with production centres distributed across many institutes worldwide. The production process is unusual in that only one final product is manufactured; however the types of parts from which it is assembled could consist of many versions. The evolution of the detector will take many years and during this process the history of versioned parts must be captured. The ultimate detector will be part of a high energy physics experiment therefore collection and storage of manufacturing & production data is just as important as control of the process. This stored data not only gives the "as built" view of the final system but has been designed as a warehouse to provide "calibration views", "maintenance views" and other views not yet conceived by the designers. It is these specialized aspects which characterize CRISTAL as a scientific workflow system. A general workflow system is used to coordinate and manage execution of the thousands of tasks and activities that occur in any complex enterprise. Workflow management can be applied to diverse applications from banking to manufacturing. In each case the system must be capable of describing and storing the tasks and activities of the domain to be automated, executing and co-ordinating the tasks and storing the outcomes. CRISTAL has taken an object-oriented approach describing all parts, manufacturing & production tasks, manufacturing & production data using meta-modeling techniques. As a consequence of the uncertainty and specialized nature of the application the core meta-model of the CRISTAL workflow system has the potential to be applied to almost any workflow or enterprise resource management application. The motivation to achieve this potential has stemmed from requests to apply the core CRISTAL technology to bio-informatics and general manufacturing domains. However a number of problems remain to be solved in order to develop our workflow software to cope with the demands of such a diverse product family range.

# 3. Software Product Lines

The product family problem is well known. A software product line [2,3] is a set of software systems that share a common set of features that satisfy a specific market demand. The key idea is to build shared assets that can be instantiated and combined to develop instances of the product line. Similar to a manufactured product line software products will:-

- Pertain to an application domain and market
- Share an architecture and
- Be built from reusable components

The application domain is reasonably clear in our particular example but the issues that surround a common architecture and components are less so. The following explore these issues in further depth based on our software engineering experiences.

## 3.1 Software Architecture

A product line software architecture[4] is the central artifact in product line engineering because it provides the framework for developing and integrating shared assets and must be common to all the products. Naturally the common user requirements map to the standard architecture but product specific requirements must map to variations provided for by the architecture. It is these specific requirement variations that define the particular product line. The problem is how to best manage and include mechanisms for this variation. [5] Discusses methods to model and capture this variation. Standard computing mechanisms to cope with variation are:-

- Alternate selection using "if then else" flow control
- Alternate selection using parameters
- And in object orientation the use of inheritance, delegation and meta-models.

In our experience in building workflow systems one of the benefits of meta-modeling is that with careful analysis and use of descriptive classes a core generic software architecture can be developed to support almost any type of workflow system. A discussion of the concepts and benefits of meta-modeling is not in the scope of this paper but more details can be found in [6].

Applying the meta-modeling approach to a product line of workflow managers (that is workflow managers for production of aircraft, cars, kitchens etc.) would necessitate describing the activities and items to configure the architecture to the particular work flow manager in the product line. Compared with a more software component based approach where the actual product line goes through a software build process where variational components are linked in or omitted according to the features of the target product. The former approach makes for a configurable adaptive architecture but the second is required to support a product line.

## 3.2 Reusable Components

Product line engineering practice advocates the generation and application of generic software assets that are reusable across a family of target products. It suggests analyzing common and variable product characteristics to define scope of reuse, identify reusable components with a suitable level of generality. The expected benefits are production of quality cost-effective software, rapid application development and improved maintenance. It emphasizes strategic planned reuse rather than opportunistic reuse. That is it is not just about libraries, class hierarchies or configurable architectures. This is reuse at a very high level disconnected from implementation issues. The following section discusses the implementation issues of software reuse and its application to components and component based development and the final part of the paper attempts to link the two together.

## 4. Software Reuse

Software reuse is not a new concept as Figure 1 illustrates. Early efforts focused on small-grained reuse of software code. Our experience over the past 10 years of building object-oriented systems has convinced us that most reuse has come from higher-level design artifacts.

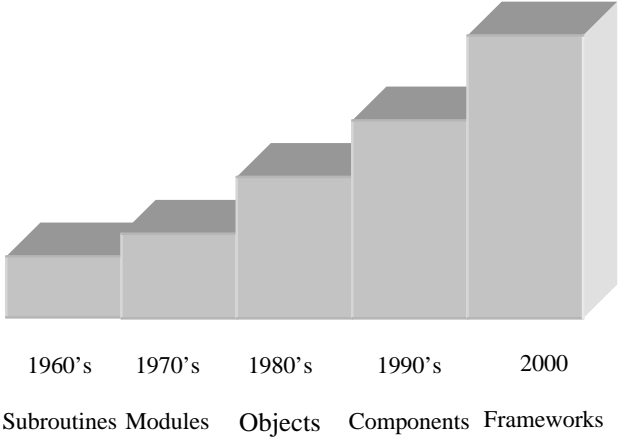

**Figure 1: A History of Software Reuse.**

Very little code has been reused, except class library reuse mainly confined to client-side user interfaces. So why so little reuse at the code level? One explanation appears to be that the cost of creation and use of these small-grained assets often outweighed the modest gains. But another important factor is that the underlying software technology is moving so fast, especially true in software projects with long time scales. For example object technology has witnessed, in a short space of time Smalltalk, ADA, C++, Java, EJB, COM+, Active X and OMG CORBA.

Where we have experienced more success in reuse of software artifacts is with visual modeling languages such as Object Modeling Technique (OMT) and the Unified Modeling Language (UML)[7]. The creation and evolution of graphical models using UML has allowed us to specify, visualize, construct and document the artifacts of the software systems we have built. Building UML models has provided a structure for problem solving and allowed us to contemplate large-scale system problems. Derived from OMT, UML version 1.1 was adopted as an Object Management Group (OMG) standard in November 1997 with a recent minor version, UML 1.3, adopted in November 1999. Usually the great thing about standards is that there are lots to choose from. However in contrast to the rapidly changing implementation software technology, UML is the universal OAD modeling standard used by OMG member organizations and Microsoft. Perhaps because of this stability we have over the years been able to reuse large-grained architectural frameworks and patterns which have been captured in UML. The term's pattern, framework, component are somewhat overloaded and the following subsections provide working definitions and discuss reuse issue experiences.

## 4.1 Patterns

A Pattern[8] is a solution schema expressed in terms of objects & classes for recurring design problems within a particular context. Patterns focus on reuse of abstract designs and software architecture, which is usually, described using graphical modeling notation. So in UML this is specification is done using interaction, class and object diagrams. The patterns that we have reused in the construction of our workflow management system[9] have evolved out of years of proven design experience. Although made up of graphical diagrams the documentation provides a vocabulary and concept understanding amongst the team. Documentation describes heuristics for use and applicability although this is not modeled in UML. In the object oriented community well known patterns are named, described and cataloged for reuse by the community as a whole. We have not only used many well-known patterns but in the domain of workflow management discovered new patterns. It has enabled us to make use of design patterns that were proven on previous projects and is a good example of reuse at the larger grain level. UML diagrams are able to describe pattern structure but provides no support for describing pattern behavior or any notation for the pattern template. UML 1.4, which is in draft stage, will enhance the notation for patterns.

## 4.2 Frameworks

A framework is the term given to a more powerful and large grained object oriented reuse technique. It is a reusable semi-complete application that can be specialized to produce custom applications [10]. It specifies a reusable architecture for all or part of a system and may include reusable classes, patterns or templates. Frameworks focus on reuse of concrete design algorithms and implementations in a particular programming language. Frameworks can be viewed as the reification of families of design patterns. When specialized for a particular application then it is called an application framework and Fayad[11] identifies three categories:

- System Infrastructure where frameworks are applied to operating systems, network communications and GUI's.
- Middleware applied to ORBs and transactions
- Enterprise Frameworks which address domains such as telecommunications, business, manufacturing.

Framework requirements are defined by software vendors or standards organizations for example IBM's San Francisco Project, FASTech' FACTORYworks, and Motorola's CIM Baseline. Fingar[12] maintains that most frameworks should capture workflows since they provide the necessary modeling capabilities for constructing any business process. He states that workflow management is one of the elements common to all e-commerce applications and is essential. Many proponents of frameworks go so far as to suggest that workflow mechanisms should eliminate the need for most application programming in the workplace.

Frameworks can also be classified according to the techniques used to extend them. Whitebox frameworks rely on OO language features such as inheritance and dynamic binding. Blackbox frameworks are structured and extended using object composition and delegation.

Component frameworks are specialized frameworks that are designed to support components. D'Souza[13] describes a component based framework as a collaboration in which all the components are specified with type models; some of them may come with their own implementations. To use the framework you plug in components that fulfill the specifications. Three main industrial examples of component frameworks are OMG's Corba Component Model (CCM) Enterprise Java Beans (EJB) and Microsoft's COM+.

## 4.3 Components

Are defined as a package of software that can be independently replaced. It both provides and requires services based on specified interfaces [13]. It conforms to architectural standards so that it can plug in and interoperate with other components. The granularity of components can vary from an instance of single to many classes and can be a significant part of a system, consistent with the goal of reuse. Unlike classes, components contain implementation elements such as source, binary executable or scripts. Components are binary-replaceable things and this distinction more than anything else sets them apart from classes. They package implementation and because of interface based design can be replaced. This means that when a new variant of a component is created it can replace a previous without recompiling other components, provided it conforms to the same interface. Software developers can build applications by assembling components rather than designing and coding.

In order to gain the payoff in software reuse as advocated by product line engineering there some difficulties to be resolved. Compared with classes and patterns which are modeled at analysis and design phase components are modeled at implementation phase. This in our experience is a particular problem where we have so much invested in graphical models. The essence of the problem is how does a collection of classes modeled at the UML OAD level, become implementation components. This raises the follow-up question; is UML capable of component modeling? Although components have become the de facto standard for desktop development this is not the case for server development. In summary leading component architectures have matured and evolved to support enterprise application. What is not clear is whether graphical modeling languages and tools can support the leading component architectures to deliver the goals of product-line engineering. The following section discusses these issues.

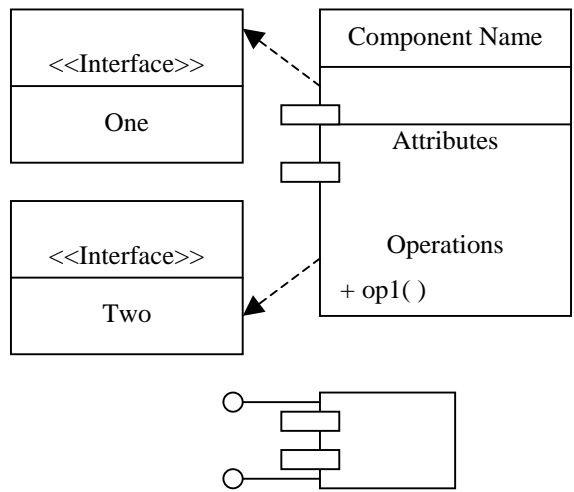

**Figure 2: UML Component Notation**

## 5. Modeling Components

A component in UML is a software artifact that exists at runtime. The notation for modeling components in UML is shown in Figure 2. In the top part of the figure the long hand notation for a component is shown complete with attributes and operations. This particular component is realized by two interfaces, interface One and interface Two. Underneath is shown the shortened notation where almost all of the detail is hidden. The two interfaces of the component are shown as so-called "lollipops". UML components are typical found in implementation related component diagrams and deployment diagrams.

The ability to model component frameworks is just as essential as being able to model components. Although component frameworks vary they do conform to a common architectural pattern. Figure 3 adapted from Kobryn[14] illustrates this common pattern using UML notation. The pattern is represented by the UML 1.4 ellipse with dashed perimeter and contains a number of classifiers. The client represents an entity that requests a service from the component. The request is never delivered directly to the component, instead it is intercepted by two proxies, FactoryProxy and RemoteProxy. The role of these entities is very important. It allows transparent insertion of common services by the component's runtime environment or container. Proxies themselves conform to well known patterns further details of which can be found in [8]. The FactoryProxy role is concerned with creation and location whilst the RemoteProxy with operations specific to the component. Both proxies and the component itself are held in a component container. The container represents the component's runtime environment and supports common distributed services such as security, transactions and persistence. Associated with each component within the container is a Context entity which stores information such as transaction and security status. The component framework pattern shows, using an XOR, that either the component or the container is responsible for managing persistence. Kobryn[14] shows that this component framework pattern applies to both EJB as well as COM+.

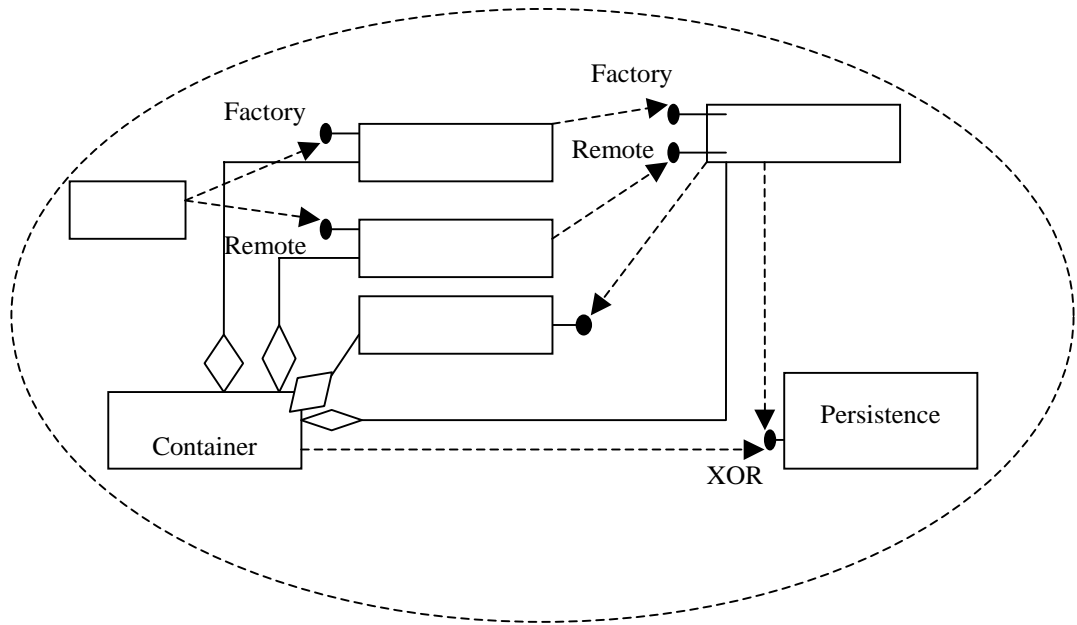

**Figure 3: A Component Framework Pattern**

Some important issues are identified when performing this comparison. Several of the classifiers in the EJB and COM + versions of the pattern are UML stereotypes, that is user-customized extensions to the language and not part of standard UML. For example EJB interfaces can declare constants whereas in UML interfaces this is not allowed. Another issue is that when combining components with various entities that realize their interfaces it is not always clear how it should take place especially in cases of complex nesting. There are no UML constructs to support large component systems and frameworks. A very important issue is the need for support to allow developers to model components earlier in the software life cycle. Several methods have been published: Catalysis (supports component modeling), Unified Process (Limited support for components), KoBra and Pulse[15] (under development but designed to support both components and product line engineering).

## 6. Conclusions

A key enabler to product line engineering is software reuse. Reusable large grain software artifacts identified at the analysis and design phase must be evolved in a consistent manner to their realizable implementation counterparts (components) at runtime. Currently there is a semantic gap between object and component-based modeling. Although the universal modeling language UML 1.3 does provide notation for components a number of major issues have been identified which restrict support for the major component technologies. The UML community is working to overcome these limitations with minor (UML 1.4) and major revisions (UML 2.0) planned. OO Frameworks promise a new vehicle for reuse but the concepts are still being evolved. So in conclusion complete life cycle component-based product engineering is not a reality, but there are signs that progress is being made.

## Acknowledgments

The authors take this opportunity to acknowledge the support of their institutes and in particular thanks to Paul Lecoq, Jean-Lious Faure, Martti Pimia and Jean-Pierre Vialle. Alain Bazan, Florida Estrella, Thierry Le Flour,

Cristoph Koch, Sopie Lieunard, Steve Murray, Giovanni Organtini, Laslo Varga, Marton Zsenei and Guy Chevenier.

# References

[1] Baker, N., McClatchey R., and LeGoff, J-M., " Scientific Workflow Management in a Distributed Production Environment" EDOC'97 Workshop Proceedings, IEEE Computer Society, 1997, pp. 291-298.

[2] Weiss, D., and Lai, C., Software Product-Line Engineering. Addison-Wesley, Reading Mass., 1999.

[3] DeBaud, J-M., and Schmid, K., "A Systematic Approach to Derive the Scope of Software Product Lines," Proc. 21st Int'l Conf. Software Eng., ACM Press, N. Y. 1999.

[4] Bayer, J., Flege, O., and Gacek, C., "Creating Product Line Architectures," Proc. 3rd Int'l Workshop Software Architectures for Product Families (IWSAPF-3), 2000

[5] Keepence, B., and Mannion, M., "Using Patterns to Model Variability in Product Families," IEEE Software, IEEE Press July/August, 1999 pp. 102 -108

[6] Barry, A., Baker, N., et al, Meta-Data Based Design of a Workflow System, OOPSLA'98: Workshop on Applications of Object Oriented Workflow Management Systems, Vancouver, Canada. October 1999.

[7] UML Revision Task Force, OMG UML, v. 1.3, document ad/99-06-08. OMG June 1999

[8] Buschmann, F., et al., Pattern-Oriented Software Architecture: A System of Patterns. Wiley, N.Y. 1996.

[9] Kovacs, Z., et al., "Patterns in a Manufacturing and Production Environment", EDOC'99 Conference Proceedings, IEEE Computer Society, 1999.

[10] Fayad, M., et al. Building Application Frameworks. Wiley, N.Y., 1999

[11] Fayad, M., "Introduction to the Computing Surveys' Electronic Symposium on Object-Oriented Application Frameworks." ACM Computing Surveys, Vol32, No 1, March 2000, pp. 1-11

[12] Fingar, P., "Component-Based Frameworks for E-Commerce," Communications of the ACM, Vol. 43, No. 10, October 2000, pp. 61-66

[13] D'Souza, D., and Wills, A. C., "Objects, Components and Frameworks with UML: The Catalysis Approach." Addison-Wesley, Reading, MA, 1999

[14] Kobryn, C., "Modeling Components and Frameworks with UML," Communications of the ACM, Vol. 43, No. 10, October 2000, pp. 31-38

[15] Bayer, J., et al., "Pulse: A Methodology to Develop Software Product Lines," Symp. Software Reusability' 99 (SSR'99), ACM Press, NY, 1999, pp. 122-131.